\documentclass[10pt,aps,prl,twocolumn,groupedaddress,superscriptaddress]{revtex4-2}

\usepackage[english]{babel}
\usepackage{amsmath} 
\usepackage{amsfonts} 
\usepackage{amssymb} 
\usepackage{braket} 
\usepackage{graphicx} 
\usepackage{mathtools}
\usepackage{xcolor} 

\definecolor{ngreen}{rgb}{0.2,0.7,0.2}

\usepackage{hyperref}

\begin{document}

\author{Tianqi Zheng}
\affiliation{State Key Laboratory for Mesoscopic Physics, School of Physics, Frontiers Science Center for Nano-optoelectronics, $\&$ Collaborative Innovation Center of Quantum Matter, Peking University, Beijing 100871, China}
\author{Yi Li}
\affiliation{State Key Laboratory for Mesoscopic Physics, School of Physics, Frontiers Science Center for Nano-optoelectronics, $\&$ Collaborative Innovation Center of Quantum Matter, Peking University, Beijing 100871, China}
\affiliation{Beijing Academy of Quantum Information Sciences, Beijing 100193, China}
\author{Yu Xiang}
\affiliation{Ministry of Education Key Laboratory for Nonequilibrium Synthesis and Modulation of Condensed Matter, Shaanxi Province Key Laboratory of Quantum Information and Quantum Optoelectronic Devices, School of Physics, Xi'an Jiaotong University, Xi'an 710049, China}
\affiliation{State Key Laboratory for Mesoscopic Physics, School of Physics, Frontiers Science Center for Nano-optoelectronics, $\&$ Collaborative Innovation Center of Quantum Matter, Peking University, Beijing 100871, China}
\author{Qiongyi He}
\email{qiongyihe@pku.edu.cn}
\affiliation{State Key Laboratory for Mesoscopic Physics, School of Physics, Frontiers Science Center for Nano-optoelectronics, $\&$ Collaborative Innovation Center of Quantum Matter, Peking University, Beijing 100871, China}
\affiliation{Collaborative Innovation Center of Extreme Optics, Shanxi University, Taiyuan, Shanxi 030006, China}
\affiliation{Hefei National Laboratory, Hefei 230088, China}

\title{Measurement-Incompatibility Constraints for Maximal Randomness}


\begin{abstract}
Certifying maximal quantum randomness without assumptions about system dimension remains a pivotal challenge for secure communication and foundational studies. Here, we introduce a generalized framework to directly certify maximal randomness from observed probability distributions across systems with arbitrary user numbers, without relying on the Bell-inequality violations. By analyzing probability distributions directly, we identify a class of quantum states and projective measurements that achieve maximal randomness in bipartite and tripartite scenarios, ensuring practical feasibility. Further analysis reveals a counterintuitive trade-off governing measurement incompatibility among users: sufficient incompatibility for one user permits arbitrarily small incompatibility for others, defying conventional symmetry assumptions in the Bell test. This asymmetry provides a pathway to optimize device-independent protocols by strategically distributing quantum resources. Our results establish a versatile and experimentally accessible route to scalable randomness certification, with implications for quantum cryptography and the physics of nonlocal correlations.
\end{abstract}

\maketitle


\textit{Introduction---} Quantum randomness serves as a fundamental resource with applications across cryptography~\cite{QKD,BeyondQKD} and scientific simulations~\cite{MC}, attracting significant attention~\cite{Pironio2010, DefOfRandom, DifferLevelOfRand2014, SteeringRandom2015, SteeringRand2022}. True random numbers can be generated using a pure source and well-characterized measurements~\cite{RevOfRand2016}, where the intrinsic randomness on outcomes is guaranteed by Born's rule. However, this approach relies on assumptions about the internal workings of the devices. To relax these assumptions, Bell nonlocality~\cite{RMP2014BellNonlocality} enables the certification of randomness in a device-independent (DI) scenario, as a violation of Bell inequality rules out the deterministic local hidden-variable model. In such a scenario, the devices can be treated as ``black boxes'' without characterizing their internal quantum states or measurements~\cite{Pironio2010, DIrandomness2014, DIQRNG, DIQRNG2, RevOfRand2016}.  

 
Substantial research efforts have been devoted to exploring the relationship between Bell inequality violations and the certified randomness. However, randomness certification through inequality violations often faces significant limitations. For instance, higher inequality violations do not necessarily imply greater randomness~\cite{TradeOffBetNonAndRand, MaxRand} and achieving the maximal violation of a Bell inequality does not always suffice for randomness certification~\cite{maxlocalnorandom}. It has been demonstrated that without the need to process the data into an inequality, the probability distributions can directly certify a higher degree of randomness~\cite{DefOfRandom}. Recent studies have pointed out that certifying maximal randomness requires ensuring the uniqueness of the probability distribution, but not all Bell inequalities exhibit unique maximal violations~\cite{MaxRand2,LiftingBell, NoUniqueBell}. Moreover, for multipartite quantum systems, the analysis of nonlocality becomes significantly complicated~\cite{GenuineNonlocal, NonlocalInNetwork1, NonlocalInNetwork2, SvetGenuNonlocal, Mermin, tripartieBellRandom}, making it more difficult to certify more randomness relying on multipartite Bell-inequality violations. 
%

Above observations motivate a direct analysis of probability distributions for certifying maximal randomness. Some studies have demonstrated instances where positive operator-valued measures (POVMs) can generate maximal randomness~\cite{MaxRandForPartialStates2020, MaxRandAndMUB, RandForPOVMEarly}. However, due to their experimental implementation challenges, developing a general and experimentally feasible approach for achieving maximal randomness remains an open research question. From a physical perspective, randomness manifests the uncertainty relations between quantum observables, suggesting that the incompatibility of the measurements employed by users should play a decisive role in randomness certification~\cite{StateAndMeas, IR3}. Recently, a necessary and sufficient connection between randomness certification and specific measurement incompatibility structures has been rigorously established~\cite{Liyi2024}. While this framework provides important insights, the quantitative relationship between the degree of randomness and the amount of measurement incompatibility remains unexplored. So several fundamental questions naturally arises: What are the essential requirements on users' measurements for achieving maximal randomness? How might we establish a quantitative relationship between the certified randomness and the underlying measurement incompatibility? 

In our work, we present a generalized method to certify maximal randomness for arbitrary numbers of users, analyzing probability distributions directly rather than employing Bell-inequality violations. We successfully apply this method to identify a series of probability distribution instances with maximal randomness and take the bipartite and tripartite scenarios as examples. Notably, these required states and measurements are simple in form and experimentally friendly. Furthermore, by analyzing the measurement incompatibilities of different users, we find that there exists a trade-off: sufficiently large incompatibility for one user permits arbitrarily small incompatibility for others--a counterintuitive relationship challenging conventional multipartite resource allocation paradigms.

\textit{Definitions and Method---}
First, we commence our analysis with the bipartite scenario, wherein two users, Alice and Bob (denoted as $A$ and $B$) are situated separately and share a quantum state $\rho_{AB}$. Both of them accept $m$ inputs, labeled by $x,y \in \{1,\dots,m\}$, and produce $n$ outputs, labeled by $a,b \in \{1,\dots,n\}$. In other words, each user can choose from $m$ different measurements, all of which have $n$ results. Given the joint probability distributions $\{P(a,b|x,y)\}$, randomness on the outcomes of measurements $x^*$ and $y^*$ can be certified through the evaluation of the guessing probability of a potential eavesdropper (Eve)~\cite{DefOfRandom}:
\begin{gather}
    P_g(x^*,y^*) = \text{max}~\sum_\lambda \max\limits_{a,b}~p_\lambda P(a,b|x^*,y^*;\lambda), \notag \\
    \text{subject to}~ P(a,b|x,y) = \sum_\lambda~p_\lambda P(a,b|x,y;\lambda), ~\forall a,b,x,y; \notag \\
    P(a,b|x,y;\lambda) \in Q,~\forall\lambda.
\end{gather}
Here $Q$ denotes the quantum set, which is known to be convex~\cite{BellNonlocality}. This property will play a crucial role in our subsequent analysis. Furthermore, the randomness is usually quantified using the min-entropy~\cite{Pironio2010}:
\begin{equation}
H(x^*,y^*) = -\log_2{P_g(x^*,y^*)}.
\end{equation}
It is straightforwardly evident that the maximum achievable randomness under these conditions is $2\log_2 n$ bits. In the following discussion, we will explore how to achieve this upper bound.

From the above definitions and the convexity of the quantum set, the following conditions for maximal randomness arise naturally~\cite{MaxRand,MaxRand2}: (1) Uniform output distribution: $P(a,b|x^*,y^*)=1/{n^2},~\forall a,b$. (2) Boundary positioning: The distribution lies on the boundary of the quantum set. The first condition stated above is self-evident, whereas the second condition admits an intuitive interpretation: By definition, Eve can optimize her guessing probability by introducing a hidden variable $\lambda$ and decomposing the observed probability distribution. However, such a decomposition is impossible when the point resides on the boundary of the convex quantum set –- a boundary which is non-flat due to the non-polyhedral nature of the quantum set.

\begin{figure}[htbp]
    \centering
    \includegraphics[width=0.4\textwidth]{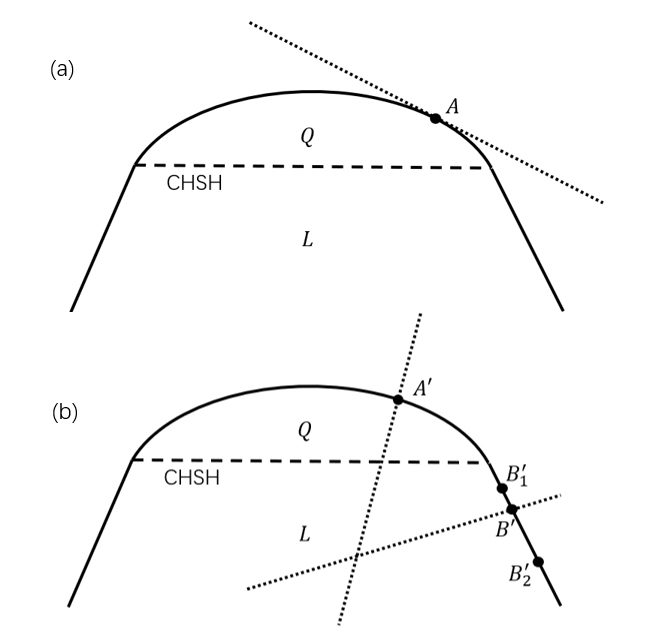}
    \caption{(a) The dashed straight line represents a linear inequality. The point of tangency ($A$) between this line and the boundary of the quantum set corresponds to the probability distribution that maximizes the violation of the inequality. (b) Other straight lines (lower-dimensional linear spaces) are constructed by introducing linear constraints. When parameterizing these lines, the positions where the parameter takes extreme values correspond to the intersection points with the boundary. If the line is generated appropriately, the intersection point will lie in the non-local region ($A'$), which is the desired point. Conversely, if the intersection point falls in the local set ($B'$), the polyhedral structure of local correlations ~\cite{BellNonlocality,MaxRand2} permits decomposition into convex combinations of other points (For example, $B_1'$ and $B_2'$), thus precluding randomness certification. }
    \label{fig:methods}
\end{figure}

We proceed from these two conditions to search for the distribution with maximal randomness, as well as the quantum states and measurements capable of generating such distributions. Previous approaches often rely on maximizing the violation of specific inequalities~\cite{MaxRand,MaxRand2}. From a geometric perspective, this method is equivalent to translating a straight line in space until it becomes tangent to the quantum set (Fig.~\ref{fig:methods}~(a)). However, this approach requires post-selection to identify tangent points that satisfy (or approximately satisfy) the uniform distribution condition. This post-selection process introduces inefficiency, making it challenging to obtain a large number of qualifying points in a single attempt. 

To address this limitation, we propose a more direct approach: optimizing (maximizing or minimizing) a specific parameter in the distribution under some linear constraints (Fig.~\ref{fig:methods}~(b)).  These constraints consist of two components: (1) the uniform distribution condition mentioned above, (2) other restrictions serving the purpose of reducing the dimensionality of the optimization problem. When the constraints are appropriately chosen, the solution to this optimization problem yields the desired probability distribution that certifies maximal randomness. Moreover, by carefully designing the form of the constraints, the optimization problem can be simplified---sometimes even admitting an analytical solution. In the following sections, we will illustrate this point through concrete examples.

\textit{Results---}
The proposed method is general and universally applicable to arbitrary $m$ and $n$, meaning it can handle quantum systems of any finite dimensions. To demonstrate this approach concretely, we consider the simplest nontrivial case where $m=n=2$. In this scenario, the mathematical formulation becomes particularly elegant and tractable. Specifically, the discussions above imply that maximal randomness can be generated by solving the following optimization problem:
\begin{gather}
    \text{min/max}~P(1,1|2,2) \notag \\
    \text{subject to}~ P(a,b|1,1) = \frac{1}{4},\forall a,b = 1,2 \notag \\
    P(1,1|1,2) = x, P(2,1|1,2) = y  \notag \\
    P(1,1|2,1) = z, P(1,2|2,1) = w \notag \\
    \label{eq:general}
    P \in Q 
\end{gather}
Here $x,y,z,w$ are four predetermined constants used to define the constraints. For later convenience, let $s\equiv z+w$, $t\equiv x+y$. These quantities have clear physical interpretations: they stand for the marginal probability of Alice and Bob, respectively, when they use their second base of measurement. The probabilities in Eq.~\eqref{eq:general} ($x,y,w,z,1/4,P(1,1|2,2)$) are enough to determine any non-signaling behavior.

\begin{figure}[thbp]
    \centering
    \includegraphics[width=0.45\textwidth]{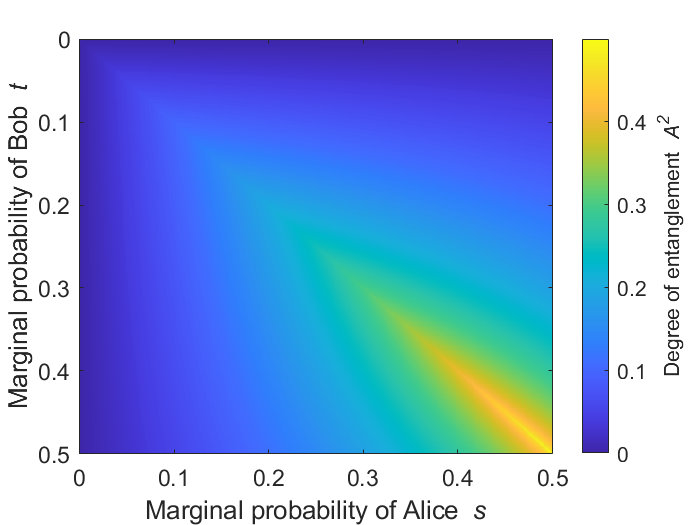}
    \caption{Entanglement of the states generating maximal randomness. The color represents the value of $A^2$ when $f(A;s,t)$ reaches the maximum. The entanglement can be arbitrarily big ($A^2\rightarrow0.5$) or arbitrarily small ($A^2\rightarrow0$).}
    \label{fig:NoMaxEntang}
\end{figure}

The optimization problem can be solved by expressing the probabilities with quantum states and measurements. Evidently, distributions located on the boundary exclusively arise from pure states~\cite{StateAndMeas}. And we only consider two-qubit states and projective measurements: $\ket{\Phi} = A\ket{00} + B\ket{11}$, $\ket{\psi}^A_{1|2} = \alpha_1\ket{0} + \beta_1\ket{1}$, $\ket{\psi}^B_{1|2} = \alpha_2\ket{0} + \beta_2\ket{1}$. Here $\ket{\psi}^A_{1|2}$ represents the projector corresponding to the first outcome of user $A$'s second measurement and the meanings of other similar symbols can be inferred by analogy. Applying Schmidt decomposition to the state $\ket{\Phi}$ and utilizing the aforementioned definitions, we derive the following relations:
\begin{gather}
    |\alpha_1|^2 = \frac{1-s-A^2}{1-2A^2},~|\beta_1|^2 = \frac{s-A^2}{1-2A^2} \notag \\
    \label{eq:aibi}
    |\alpha_2|^2 = \frac{1-t-A^2}{1-2A^2},~|\beta_2|^2 = \frac{t-A^2}{1-2A^2}
\end{gather}
Consequently, the objective function $P(1,1|2,2)$ admits the upper bound:
\begin{equation}
\label{eq:OpWithA}
\begin{aligned}
    P(1,1|2,2) &=  |A\alpha_1\alpha_2+B\beta_1\beta_2|^2 \leq \max_{A}f(A;s,t)
\end{aligned}
\end{equation}
where $f(A;s,t) \coloneqq \frac{1}{1-2A^2}[\sqrt{A^2(1-s-A^2)(1-t-A^2)} 
+\sqrt{(1-A^2)(s-A^2)(t-A^2)}]^2$ and the derivation process uses the Cauchy–Schwarz inequality. Here we take the expression $f(A;s,t)$ as a function of $A$ with fixed value of $s$ and $t$. Without loss of generality, $s$, $t$ and $A^2$ are constrained to the range 0 to 0.5 to ensure that no negative numbers appear under the square root in the formula. In principle, this one-dimensional optimization problem admits an analytical solution through direct derivation. The corresponding numerical results are presented in Fig.~\ref{fig:NoMaxEntang}. Notably, the results demonstrates that any arbitrary entangled pure state suffices to certify maximal randomness. 

To achieve maximal randomness with these general states, the measurement operators must be carefully chosen. Specifically, users need to solve the optimization problem in Eq.~\eqref{eq:OpWithA} to determine both the magnitudes and relative phases of $\alpha_1$, $\beta_1$, $\alpha_2$, $\beta_2$, while simultaneously satisfying additional constraints that guarantee the uniform output distribution condition. For computational tractability, in the following discussions we restrict our analysis to a special case where users employ maximally entangled quantum states. This simplification allows us to examine the properties of the measurements employed by the users, with particular emphasis on quantifying the incompatibility between different measurements used by the same user. Randomness generation should stem from the uncertainty principle inherent in quantum mechanics. This naturally leads to the conjecture that achieving maximal randomness requires highly incompatible measurements. To quantify this relationship, we adopt the well-established framework of incompatibility robustness~\cite{IR, IRRev, IR3}, which provides a rigorous metric for characterizing the degree of measurement incompatibility:
\begin{gather}
    \text{max}~\eta, \notag \\
    \text{s.t.}~~M^\eta_{a|x}=\eta M_{a|x} + (1-\eta)\frac{I}{2}; \notag \\ M^\eta_{a|x} \text{are compatible}.
\end{gather}

For clarity of demonstration, we revisit Eq.~\eqref{eq:aibi}: When the parameters $s$ and $t$ are fixed at $1/2$, Eq.~\eqref{eq:aibi} shows that if $A^2\neq\frac{1}{2}$, the right-hand expression simplifies trivially by cancellation, just fixing the measurement parameters, which is not our focus. We therefore set $A^2=\frac{1}{2}$, rendering Eq. (4) indeterminate. Further analysis is then required to extract measurement information. Specifically, let $\ket{\Phi}=\frac{1}{\sqrt{2}}(\ket{00}+\ket{11)}$. The first set of measurements are selected as $\sigma_x$ and $\sigma_z$, for Alice and Bob respectively, to satisfy the uniform distribution condition. Then, it can be derived that $|\alpha_2 + \beta_2|^2 = 4x$, $|\alpha_1|^2 = 2z$, and $|\alpha_1\alpha_2+\beta_1\beta_2|^2 = 2P(1,1|2,2)$. Taking use of Cauchy–Schwarz inequality again, the minimum value of the objective function $P(1,1|2,2)$ can be determined as follows:
\begin{equation}
\label{eq:inf}
\begin{aligned}
    \sqrt{2P(1,1|2,2)} = |\alpha_1\alpha_2 + \beta_1\beta_2| \leq g(x,z),
\end{aligned}
\end{equation}
where we introduce the notation for convenience: $g(x,z) \coloneqq \frac{1}{\sqrt{2}}[\sqrt{2z}(\sqrt{2x}-\sqrt{1-2x})-\sqrt{1-2z}(\sqrt{2x}+\sqrt{1-2x})]$. Note that $g(x,z)$ is symmetric with respect to $x$ and $z$. The derivation of the final expression relies on the assumptions: $g(x,z) \geq 0$, and $x,z \leq 0.5$.

\begin{figure}[thbp]
    \centering
    \includegraphics[width=0.45\textwidth]{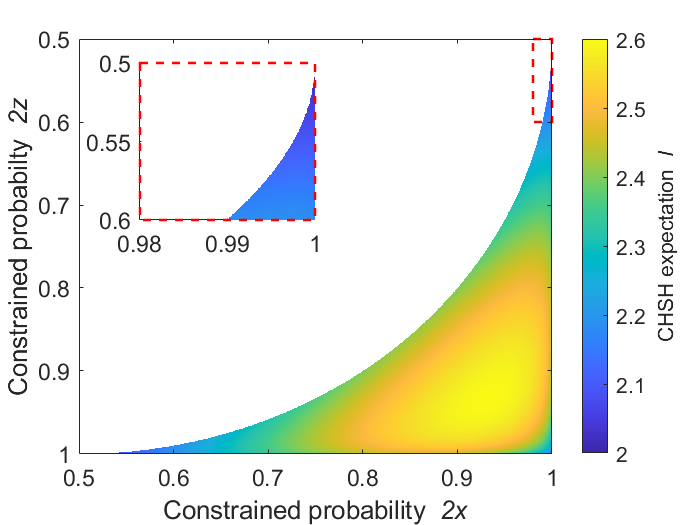}
    \caption{The values of CHSH expression $I =\langle A_1 B_1 \rangle + \langle A_1 B_2 \rangle + \langle A_2 B_1 \rangle - \langle A_2 B_2 \rangle$ are illustrated, for different $x$ and $z$. The blank area results from the failure to satisfy the constraints of $x$, $z$ and $g(x, z)$. As $2x\rightarrow1$ and $2z\rightarrow0.5$ (or vice versa), the violation of inequality~($I-2$) tends to 0 but never reaches 0, which ensures the nonlocality of this group of distributions.}
    \label{fig:CHSH}
\end{figure}

\begin{figure}[t]
    \centering
    \includegraphics[width=0.42\textwidth]{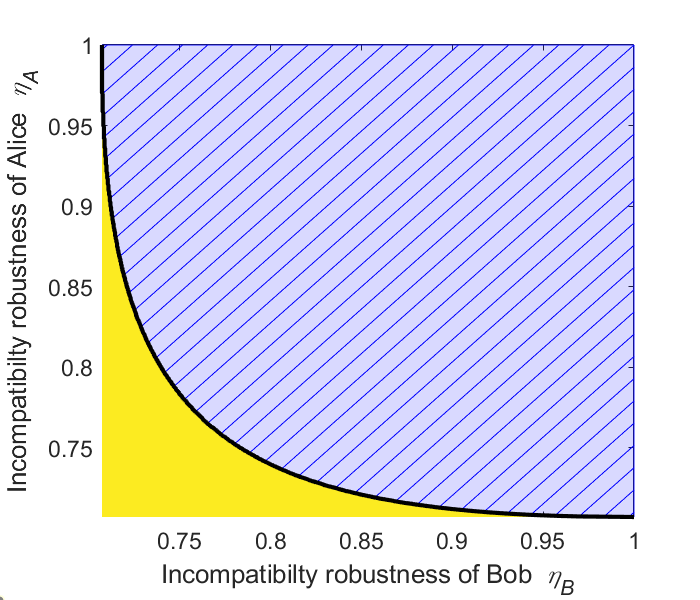}
    \caption{The constrained relationship between $\eta_A$ and $\eta_B$. The yellow region bounded below the specified line permits the generation of maximal randomness.}
    \label{fig:trade}
\end{figure}

We have successfully derived a set of quantum-boundary probability distributions by solving the optimization problem in Eq.~\eqref{eq:general}. Their nonlocality guarantees the generation of maximal randomness, as shown in Fig.~\ref{fig:CHSH}. Furthermore, based on the preceding analysis, the measurement operators employed by users $A$ and $B$ are directly determined, specifically the magnitudes and relative phases of the parameters $\alpha_i$ and $\beta_i$. This enables us to examine the incompatibility between the two measurement settings for each user. The requirement for incompatibility robustness is illustrated in Fig.~\ref{fig:trade}: To achieve maximal randomness, qualitatively, both $A$ and $B$ must employ incompatible measurements; quantitatively, there exists a trade-off between their degrees of incompatibility. sufficiently large incompatibility for one user ($\eta \rightarrow \frac{\sqrt{2}}{2}$) permits arbitrarily small incompatibility for others ($\eta \rightarrow 1$), which greatly relaxes the requirements for users to choose measurements.

The aforementioned analysis specifically addresses the two-user scenario. However, in principle, our proposed methodology can be generalized to systems with an arbitrary number of users. In the multipartite scenario, globally maximal randomness ($p\log_2 n$ bits for $p$ users) remains attainable through the formulation and solution of an optimization problem. More concretely, one can generalize Eq.~\eqref{eq:general} to the multipartite case by modifying the probabilities in the uniform distribution constraint to $1/n^p$, parameterizing the distributions $P$, and optimizing one of them. Given the growing research interest in large-scale, multi-user quantum networks~\cite{Network2011, Network2017, Network2021PRL, NetRand2024}, we extend the previous discussion on measurement incompatibility to a three-user scenario, where the participants are denoted as $A'$, $B'$ and $C'$~\cite{2018q,LiYi2023}. 

To address the complexity arising from the numerous parameters in tripartite probability distributions, we implement two simplifications: (1) Only considering maximally entangled states (GHZ states) (2) Assuming that the first two users, $A'$ and $B'$, are wholly identical. This reduction renders the tripartite problem analogous to the bipartite case analyzed before and two parameters, $x$ and $z$ are introduced likely in the constraints:$P(1,1,1|1,1,2) = x,~P(1,1,1|1,2,1) = P(1,1,1|2,1,1) = z$. The complete optimization problem can be found in the supplementary materials. It turns out that the lower bound of the objective function $P(1,1,1|2,1,2)$ is characterized by an expression analogous to Eq.~\eqref{eq:inf}:
\begin{equation}
    \label{eq:SolToTri}
    \sqrt{4P(1,1,1|2,1,2)} \geq  g_T(x,z)
\end{equation}
Here $g_T(x,z)$ exhibits a form analogous to $g(x,z)$, differing only by the substitution $2x\rightarrow4x,~2z\rightarrow4z$: $g(x,z) \coloneqq \frac{1}{\sqrt{2}}[\sqrt{4z}(\sqrt{4x}-\sqrt{1-4x})-\sqrt{1-4z}(\sqrt{4x}+\sqrt{1-4x})]$. Similarly, this result can be derived using the same method as in Eq.~\eqref{eq:inf}. While the choice of maximally entangled states provides formal consistency with the bipartite case, we emphasize that this approach lacks rigorous justification. Unlike bipartite systems where Schmidt decomposition validates the use of maximally entangled states, tripartite systems generally do not admit such decomposition~\cite{TriSchmidt}. Consequently, other quantum states might yield smaller values of the objective function, and the correctness of Eq.~\eqref{eq:SolToTri} must be tested numerically. Our numerical analysis confirms the validity of this solution, demonstrating a relative error below $10^{-5}$. Complete numerical results are provided in the supplementary materials. 

Finally, the optimization problem in the tripartite scenario admits a solution analogous to the bipartite case, with the optimized distributions generated using the same measurement configurations. The incompatibility of these measurements remains consistent with the trends illustrated in Fig.~\ref{fig:trade}, provided the substitutions $A\rightarrow A'$, $B\rightarrow C'$ are applied. Our analysis reveals that globally maximal randomness is achievable as long as the measurement incompatibility of one party ($C'$) is sufficiently large. 

\textit{Conclusion---}
We present a general framework for identifying probability distributions capable of certifying maximal randomness, applicable to quantum systems with arbitrary numbers of users and finite dimensions. Through concrete  examples in both bipartite and tripartite cases, we demonstrate that global maximal randomness can be reliably achieved when as long as one participant employs measurements with sufficiently large incompatibility. This finding suggests strong potential for extension to more complex multipartite systems and broader operational scenarios.

\textit{Acknowledgments---}
This work was supported by Beijing Natural Science Foundation (Grant No.~Z240007, NO.~QY24010), and the National Natural Science Foundation of China (No.~12125402, No~12350006, No~123B2073), the Innovation Program for Quantum Science and Technology (No.~2021ZD0301500). 
Y. X. acknowledges the support received by the National Cryptologic Science Fund of China (Grant No.~2025NCSF02048).

\bibliography{main}

\end{document}